\documentclass{eas}
\usepackage{graphicx}
\usepackage{amssymb}
\usepackage{rotating}

\usepackage[sectionbib]{natbib}
\bibpunct{(}{)}{;}{a}{}{,}
\begin{document}


\title{Detection and Characterization of Planets in Binary and Multiple  
Systems}\thanks{I wish to thank the conference organizers for their invitation
and for their warm welcome. I acknowledge the contribution of S. Udry, 
M. Mayor, J.-L. Beuzit, G. Chauvin, A.-M. Lagrange, 
T. Mazeh, and Y. Segal to the work described in Sects.~\ref{imaging} and
\ref{pib}. I acknowledge support from the French ANR   
through project grant ANR-NT-05-4\_44463 and from the 
PNP of CNRS/INSU.}
\runningtitle{A. Eggenberger: Observation of planets in binaries}
\author{A. Eggenberger}
\address{Laboratoire d'Astrophysique de Grenoble, 
UMR 5571 CNRS/Universit\'e J. Fourier, BP~53, F-38041 Grenoble Cedex 9, 
France; \email{Anne.Eggenberger@obs.ujf-grenoble.fr}}
\begin{abstract}
Moderately close binaries are a special class of targets for planet 
searches. From a theoretical standpoint, their hospitality to giant planets 
is uncertain and debated. From an observational standpoint, many of these 
systems present technical difficulties for precise radial-velocity measurements 
and classical Doppler surveys avoid them accordingly. In spite of these
adverse factors, present data support the idea that giant planets residing in 
binary and hierarchical systems provide unique observational constraints on 
the processes of planet formation and evolution. The interest and 
the importance of including various types of binary stars in extrasolar planet 
studies have thus grown over time and significant efforts have recently 
been put into: (i) searching for stellar companions to the known planet-host 
stars using direct imaging, and (ii) extending Doppler planet searches to 
spectroscopic and moderately close visual binaries. In this contribution we 
review the observational progresses made over the past years to detect and
study extrasolar planets in binary systems, putting special emphasis on the 
two developments mentioned above.
\end{abstract}
\maketitle

\section{Introduction}
\label{intro}

Nearby G-K dwarfs, which are first-choice targets for many planet search 
programs, are more often found in binary or multiple systems than in 
isolation \citep[e.g.][]{Duquennoy91,Eggenberger04b}. 
This observation, coupled with the growing evidence that many young 
binaries possess circumstellar or circumbinary disks susceptible of
sustaining planet formation \citep[e.g.][]{Monin07}, 
raises the question of the existence of planets in 
star systems of different types.

From a dynamical standpoint, planets residing in binary systems can be found in 
three different configurations: (1) in circumstellar orbits (i.e. orbiting the 
primary or the secondary star, also called S-type orbits); (2) in circumbinary 
orbits (i.e. circling both stars, P-type orbits); or (3) around the L4 or L5 
Lagrange point in systems with a very small mass ratio (Trojan planets, 
L-type orbits). However, the presence of an additional stellar companion may 
threaten either the formation or the long-term stability of planetary systems, 
imposing further restrictions. For instance, a stellar companion within 
$\sim$100 AU will likely affect -- and possibly inhibit -- the formation of 
circumstellar giant planets (e.g. \citeauthor{Nelson00}~\citeyear{Nelson00}; 
\citeauthor{Mayer05}~\citeyear{Mayer05}; 
\citeauthor{Thebault06}~\citeyear{Thebault06}; see also the
contributions by Kley, Marzari, and Th\'ebault), while a more distant but 
highly inclined companion can influence the evolution of these planets 
on secular timescales \citep[e.g.][]{Innanen97,Takeda05}. These considerations 
raise two fundamental questions: What types of binary systems do actually host 
planets in S/P/L-type orbits?, and Are such planets common or rare?

Most of the information we presently have about planets in binaries comes 
from ``classical'' Doppler planet searches which target nearby G-K dwarfs.
This observational material is highly incomplete with respect to the 
closest binaries, however, because it is difficult to extract precise radial velocities 
when the two components simultaneously contribute to the recorded flux 
\citep{EggenbergerUdry07}. To avoid light contamination at the 
spectrograph entrance, Doppler surveys exclude from their target 
samples most -- but usually not all -- double stars closer than 
$2$-$6^{\prime\prime}$ \citep[e.g.][]{Udry00,Perrier03,Marcy05,Jones06}, 
systems which we will call moderately close binaries. This  
strategy implies that classical Doppler programs provide little 
information about the existence of planets in spectroscopic and visual 
binaries $\lesssim$200~AU. To probe the occurrence of planets in these 
moderately close systems, new methods and alternative detection techniques 
have been actively developed over the past years.

This chapter is partly a review on the observational efforts to detect and
characterize 
extrasolar planets in binary systems, and partly a description of our own 
contribution to this research field. In Sect.~\ref{obs} we present a brief 
overview of the methods that are being used to detect extrasolar planets in
binary and multiple systems. In Sect.~\ref{classical_doppler} we summarize the 
information gathered on planets in binaries through classical Doppler planet 
searches. We then describe our recent efforts to investigate 
the impact of stellar duplicity on the occurrence 
and properties of giant planets. This work follows two complementary 
approaches: searching for stellar companions to the known planet-host stars 
using direct imaging (Sect.~\ref{imaging}), and extending Doppler planet 
searches to spectroscopic binaries (Sect.~\ref{pib}). We conclude in 
Sect.~\ref{conclusion} with a summary of the main results and future 
perspectives.

\section{Observational methods to detect extrasolar planets in binary systems}
\label{obs}

The interest for planets in binaries increased rapidly in $\sim$2002 following 
two important discoveries. Firstly, two giant planets were detected in 
the spectroscopic binaries GJ\,86 \citep{Queloz00} and $\gamma$\,Cephei 
\citep{Hatzes03}, bringing evidence that circumstellar Jovian planets do 
exist in systems separated by $\sim$20 AU. Secondly, \citet{Zucker02} 
pointed out that planets found in binaries seem to follow a different period-mass 
correlation than planets orbiting single stars. This led to the idea
that planets found in binary and multiple systems may provide unique 
testing grounds for the models of planet formation and evolution. 

Since $\sim$2002, a few Doppler planet searches targeting exclusively
spectroscopic and visual binaries have been initiated to complement the 
observations from classical surveys (see also Sect.~\ref{pib}, 
and the contributions by Konacki and Desidera). Programs searching for 
circumstellar planets in visual binaries $\gtrsim$100 AU 
\citep{Desidera07b,Toyota09} treat their targets as two single stars and face 
little technical difficulty. An attractive aspect of these programs is their
ability to probe possible differences in the chemical composition of solar-type
stars with and without planetary systems. Doppler surveys dedicated to planet 
searches in spectroscopic and moderately close visual binaries 
\citep{Eggenberger03,Konacki05b,Toyota05,EggenbergerUdry07} are much 
more challenging technically (the two stars cannot be observed individually), 
but can potentially bring fundamental information for planet formation 
theories. Using the same technique but different target samples, these surveys 
can search either for circumstellar planets in binaries $\lesssim$50 AU, or 
for circumbinary planets around close binaries ($\sim$0.05-0.5 AU).
None of these dedicated Doppler programs has detected a reliable planet 
candidate so far, but all the surveys are still ongoing.

Like Doppler spectroscopy, transit photometry works normally with regard to 
planet searches around the components of wide binaries, but faces additional 
challenges with moderately close systems. Yet, eclipsing binaries represent 
an attractive class of targets for the photometric method. On the one hand  
planetary transits are more likely to occur is these edge-on systems, and 
high-precision photometry should allow the detection of transiting 
circumstellar, circumbinary \citep{Deeg98,Doyle00,Ofir08}, and Trojan planets 
\citep{Caton00}. On the other hand, the same photometric data can be used to 
search for nontransiting giant planets in circumbinary 
orbits through the precise timing of eclipse minima \citep{Deeg00,Deeg08,Lee09}. 
To date, planet searches in eclipsing binaries have detected a pair of 
circumbinary substellar objects (minimum masses of 19.2 and 8.5 M$_{\rm Jup}$) 
around HW Virginis \citep{Lee09} and a possible circumbinary planet around 
CM Draconis \citep{Deeg08}.  

Given a sufficient time baseline, pulse timing measurements provide a good 
dynamical description of nearly any type of multiple system orbiting a neutron
star that can be timed with a microsecond precision. For instance, the series 
of timing data of PSR B1620-26 indicates that this millisecond pulsar belongs 
to a hierarchical triple system, with a circumbinary planet orbiting the inner 
pulsar\,--\,white dwarf pair \citep{Thorsett99,Ford00b,Sigurdsson04}. Pulsar 
timing and photometry of eclipsing binaries are currently the 
only two methods that can detect low-mass planets in/around binary systems.

According to recent modeling work, the signatures of both a planet and a 
binary companion can be detected under certain conditions in the light curves 
of high-magnification microlensing events. In particular, the microlensing 
technique should be able to identify circumstellar giant planets in binary 
systems $\lesssim$100 AU \citep{Lee08}, or Jupiter-mass circumbinary planets 
orbiting binaries separated by $\sim$0.15-0.5 AU \citep{Han08}. In the future, 
microlensing searches may thus enrich the samples of planets residing in and 
around (moderately) close binaries.

Astrometry is intrinsically well suited to search for planets in some types of 
moderately close binaries, the secondary star providing a convenient positional 
reference. Relative astrometry has the advantage of yielding the full planetary 
orbit and the planet's true mass, but the disadvantage of not distinguishing 
between circumprimary and circumsecondary planets. Since 2003 the PHASES 
program has used the phase-referencing technique at the Palomar Testbed 
Interferometer to search for circumstellar giant planets among $\sim$50 
binaries with a median separation of 19 AU 
\citep{Lane04,Muterspaugh06a}. Contrary to Doppler spectroscopy, astrometry is 
mostly sensitive to long-period companions, and the two methods nicely 
complement to probe the occurrence of giant planets in moderately close 
binaries. Current results from the PHASES program exclude the presence of giant 
planets in 8 systems from its target list \citep{Muterspaugh06b}. 

Narrow-field, adaptive optics imaging may be an alternative means to perform relative 
astrometry of moderately close binaries with a precision good enough to detect 
circumstellar massive planets (see the contributions by He{\l}miniak and Roell). 
Since this method is particularly well suited to study binary systems separated 
by a few arc seconds, it would offer the possibility of deriving 
the true mass of some of the giant planets detected by classical Doppler 
surveys. Although encouraging results have recently been reported for 
short-term observations \citep{Neuhauser07,Helminiak08,Roell08}, the technical 
and practical feasibility of this approach remains to be demonstrated in the 
long term.


\section{Results from classical Doppler planet searches}
\label{classical_doppler}
 
\subsection{The sample of planets in binaries}
\label{pib_sample}

Thanks to classical Doppler surveys and to complementary searches 
for common proper motion companions to planet-host stars (Sect.~\ref{imaging}), 
the number of planets known to orbit a component of a binary or multiple star 
has grown rapidly in the past few years and now outnumbers 53 planets in 45 
planetary systems \citep[e.g.][]{EggenbergerUdry07}. Most of these planets 
are gas giants ($>$0.3 M$_{\rm Jup}$), which mainly reflects the high 
sensitivity of the Doppler technique to massive companions. In terms of system 
architecture, these planets reside in binary or hierarchical systems with 
projected separations between $\sim$20~AU and $\sim$12000~AU, and almost 
all of them orbit the primary component. This last feature is partly a 
selection effect, the secondaries being often too faint to belong to the target 
samples used by Doppler programs. Due to their discrimination against the 
closest binaries, classical programs do not provide us with much information 
about the existence of circumbinary planets. Nonetheless, one candidate may 
have been found \citep{Correia05}.

Only a handful of the known planets reside in binaries $\lesssim$100~AU, making 
giant planets apparently rarer in these systems than in wider pairs or around
single stars. Although some theoretical models predict a shortage of giant 
planets precisely in binaries $\lesssim$100~AU 
\citep{Nelson00,Mayer05,Thebault06}, the small number of planetary detections 
in these systems results at least partly from the selection effects mentioned in
Sect.~\ref{intro}. Another interesting feature visible in the data from Doppler
surveys is the lack of circumstellar planets in binaries $\lesssim$\,20~AU. 
According to theoretical models the formation of giant planets is severely 
hampered in these systems \citep{Nelson00,Thebault04,Mayer05,Boss06,Thebault06}, 
which suggests that the ``limit'' at 20 AU might have a true meaning. 
Yet, the present observational material does not allow us to rule out 
the alternative hypothesis that the lack of planetary detections in systems 
$\lesssim$20~AU simply reflects the discrimination against ``short-period'' 
($\lesssim$10 years) spectroscopic binaries. 

To sum up, classical Doppler surveys have brought evidence that at least 
17\% of the known planetary systems are associated with one or more stellar 
companion. However, due to noticeable selection effects against moderately 
close binaries, the data from these surveys cannot be used to derive the 
true frequency of planets in binaries $\lesssim$100~AU, nor to probe the 
existence of planets in binaries $\lesssim$20~AU. To investigate 
these two fundamental questions we need planet search programs capable of 
dealing with moderately close binaries. We discuss our own Doppler surveys and 
some of their preliminary results in Sect.~\ref{pib}.

\subsection{Different properties for planets in binaries?}
\label{diff_prop}

\begin{figure}
\centering
\resizebox{\textwidth}{!}{
\includegraphics{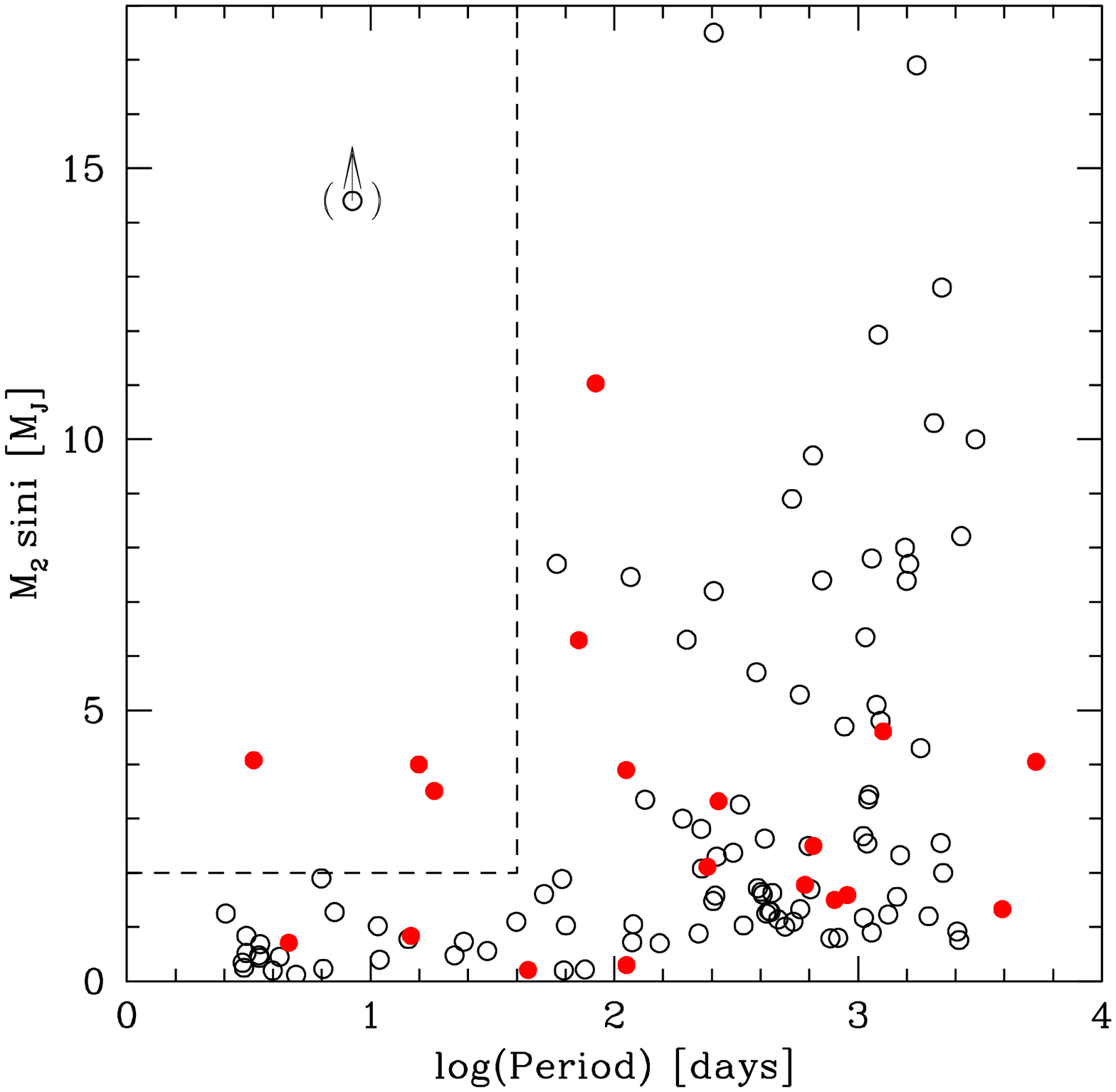}\hspace{0.5cm}
\includegraphics{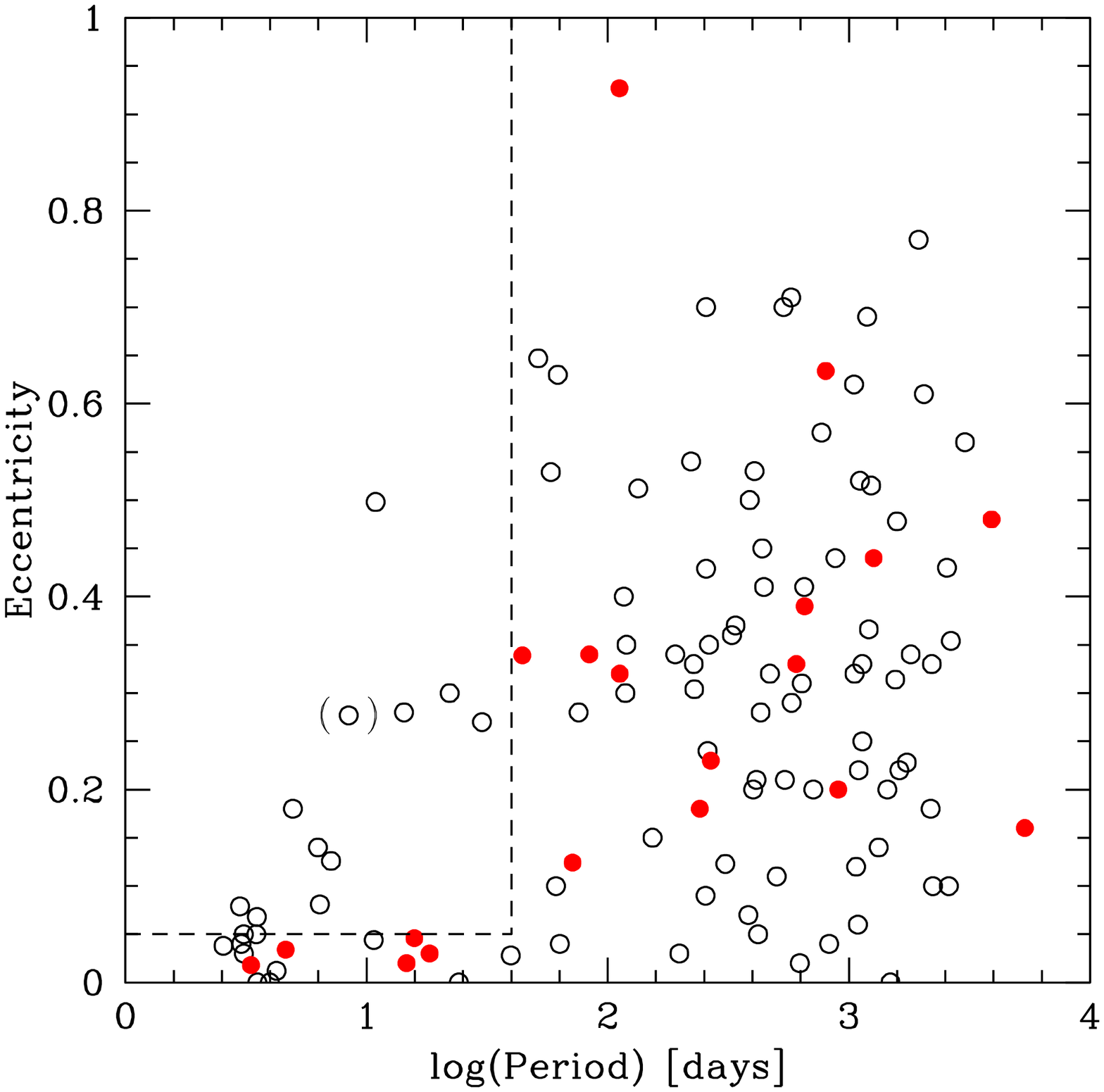}}
\caption{{\bf Left:} Minimum mass vs orbital period for all the extrasolar 
planetary candidates known in 2004. Planets orbiting a single star are 
represented as open circles, while planets residing in binary or multiple 
systems are represented as dots. The dashed line approximately
delimits the zone where only extrasolar planets belonging to binaries are found.
{\bf Right:} Eccentricity vs orbital period for the same planetary
candidates as before. The dashed line approximately delimits the region  
where no planet-in-binary is found.}
\label{stat3}
\end{figure}

Following \citet{Zucker02}, who pointed out that planets in binaries seem 
to follow a different period-mass correlation than planets orbiting single 
stars, we performed in 2004 a statistical study considering both the 
period-mass and the period-eccentricity diagrams \citep{Eggenberger04}. 
As shown in Fig.~\ref{stat3} (left), our analysis confirmed that the few most 
massive ($M_2\sin{i}$\,$\gtrsim$\,$2$~M$_{\rm Jup}$) short-period 
($P$\,$\lesssim$\,$40$~days) planets all orbit a component of a binary or 
multiple star. However, the inclusion of several new planets in binaries with 
periods $>$100~days and minimum masses in the range 
3--5~M$_{\rm Jup}$ decreased the significance of the negative period-mass 
correlation found by \citet{Zucker02}. More recent studies confirm this 
trend \citep{Desidera07,Mugrauer07}, leaving as a robust feature only the
observation that the few most massive short-period planets are all found in
binary or multiple systems.

Regarding the period-eccentricity diagram, our analysis emphasized that the 
planets with a period $P$\,$\lesssim$\,$40$~days and residing in binaries 
tend to have low eccentricities ($e$\,$\lesssim$\,$0.05$) compared to their
counterpart orbiting single stars (Fig.~\ref{stat3}, right). 
The confirmation -- or refutation -- of this trend 
looks more tricky \citep{Desidera07,Mugrauer07}, probably because several 
different mechanisms play a role in shaping the 
eccentricity distribution of extrasolar planets. We plan to revisit this 
question once we will have in hands the final results from 
our two imaging programs (Sect.~\ref{imaging}). 

Another intriguing feature is the observation that the 
four planets with the highest eccentricities ($e>0.8$) all have a stellar or 
brown dwarf companion \citep{Tamuz08}. This association likely points towards  
eccentricity excitation by the Kozai mechanism \citep{Wu03,Takeda05,Moutou09}. 
The Kozai mechanism is a secular interaction which operates in hierarchical
triple systems with high relative inclination, causing large-amplitude 
periodic oscillations of the eccentricity of the inner pair (e.g.
\citeauthor{Holman97}~\citeyear{Holman97}; 
\citeauthor{Innanen97}~\citeyear{Innanen97}; 
\citeauthor{Mazeh97}~\citeyear{Mazeh97}; 
\citeauthor{Ford00}~\citeyear{Ford00}). If distant stellar companions 
commonly induce Kozai oscillations in planetary systems, this should produce a 
small excess of very eccentric orbits among the population of planets in 
binaries \citep{Takeda05}, which is precisely what \citet{Tamuz08} pointed out.

The three trends outlined above are very interesting because they may shed 
light on the origin of some of the short- and intermediate-period planets. 
Recent theoretical work has shown that the coupling of Kozai oscillations with 
tidal friction (sometimes called Kozai migration) can produce eccentric planets on intermediate 
orbits \citep{Wu03}, but also short-period planets on (nearly) circular orbits 
such as the hot Jupiters \citep{Fabrycky07,Wu07}. Interestingly, this migration 
mechanism specific to multi-body systems could be more effective than type II 
migration in bringing massive planets to the vicinity of their host star. 
Kozai migration triggered by a distant stellar companion may thus explain why 
the most massive short-period planets are all found in binary or multiple 
systems \citep{Takeda06,Fabrycky07}. According to some 
authors, Kozai migration might also circularize planetary orbits to greater 
orbital periods \citep{Fabrycky07}, thereby explaining the trend 
seen in the eccentricity-period diagram. 

To put the above observational results on firmer 
statistical grounds, future investigations will have to improve on three 
points: (1) to enlarge the sample of short-period planets found in 
binaries, (2) to systematically probe the presence of stellar companions to 
planet-host stars, and (3) to correct for the 
selection effects against moderately close binaries. 
We describe in the next two sections our efforts to tackle these issues  
with the aim of better understanding the impact of stellar duplicity on the 
occurrence and properties of giant planets.


\section{Results from our imaging surveys}
\label{imaging}

Theoretical work indicates that the main issue regarding the occurrence of 
circumstellar giant planets in binaries $\lesssim$100 AU is whether these planets can form in 
the first place. Interestingly, the two mechanisms susceptible of forming giant 
planets -- core accretion and disk instability -- may exhibit a different 
sensitivity to the presence of a moderately close stellar companion 
\citep{Nelson00,Mayer05,Boss06,Thebault06}. In
particular, it has been suggested that giant planets formed by disk instability 
should be rare in binaries separated by 60-100 AU, while giant planets formed 
by core accretion should be common in these same systems  
\citep{Mayer05}. In practice, dynamical interactions may complicate this 
simple picture by depositing a few giant planets in systems originally void of 
any \citep{Pfahl05,PortegiesZwart05,Pfahl06}. 
Nonetheless, quantifying the occurrence of circumstellar giant planets 
in binaries $\lesssim$100~AU and studying how this occurrence varies with 
binary separation is fundamental to understanding planet formation.

As mentioned previously, the data from classical Doppler surveys cannot be used 
to derive the true frequency of planets in binaries $\lesssim$100~AU.
However, the problem of quantifying the impact of stellar duplicity on 
planet occurrence can be tackled in an indirect way, by comparing the 
multiplicity among planet-host stars to the multiplicity among similar stars 
but without planetary companions. Indeed, if the presence of a nearby stellar 
companion hinders (favors) planet formation, the frequency of circumstellar 
planets in binaries closer than a given separation -- modulo eccentricity and mass 
ratio -- should be lower (higher) than the nominal frequency of planets around 
single stars. That is, the binary fraction among planet-host stars should be 
smaller (larger) than the binary fraction among single stars. Note that such a 
comparison requires the use of a well-defined control sample to: (i) take into account the selection 
effects against moderately close binaries in Doppler planet searches, and (ii) 
compare the multiplicity among planet-host stars with the multiplicity among 
similar stars but without planetary companions. 

Since 2002 direct imaging has been used by several groups to detect new stellar 
companions to planet-host stars
(\citeauthor{Luhman02}~\citeyear{Luhman02};
\citeauthor{Patience02}~\citeyear{Patience02};
\citeauthor{Chauvin06}~\citeyear{Chauvin06}; 
\citeauthor{Mugrauer07}~\citeyear{Mugrauer07}; 
see also the contribution by Mugrauer). 
In addition, astronomical catalogs and multiepoch images from the STScI 
Digitized Sky Survey have been searched for unrecognized stellar companions to 
the known planet-host stars \citep{Raghavan06,Desidera07}. While these surveys 
have revealed many new binary and multiple systems among planet-host 
stars, they cannot draw reliable conclusions about the impact of stellar 
duplicity on planet occurrence because they lack a well-defined control 
sample of non-planet-bearing stars. 

To test whether the frequency of circumstellar giant planets is reduced 
in binaries $\lesssim$100 AU, we have conducted a 
large-scale adaptive optics search for stellar companions to $\sim$200 
solar-type stars with and without planets \citep{Eggenberger07b}. 
Our main program is divided into 
two subprograms: a southern survey (130 stars) carried out with NAOS-CONICA 
(NACO) on the Very Large Telescope, and a northern survey ($\sim$70 stars) 
carried out with PUEO on the Canada-France-Hawaii Telescope. The NACO 
survey is now completed, while the PUEO survey is still halfway. We discuss
below the results from our NACO survey.

\subsection{The NACO survey}

The NACO survey relies on a subsample of 57 planet-host stars, and on a control 
subsample of 73 dwarfs from the CORALIE planet search program showing no 
obvious evidence for planetary companions from radial-velocity measurements. 
Note that selecting the control stars within the CORALIE sample 
naturally ensures that we match the target selection criteria for Doppler 
planet searches. To avoid duplicating existing observations, we excluded  
from our survey most of the planet-host stars observed by 
\citet{Patience02} and \citet{Chauvin06}. These stars are included in our  
statistical analysis, though, which finally balances the two subsample 
sizes to $\sim$70 stars each. 

Our NACO data revealed 95 companion candidates in the vicinity of 33 targets. 
On the basis of two-epoch astrometry we identified 19 true companions, 
2 likely bound objects, and 34 background stars \citep{Eggenberger07b}. 
Among planet-host stars, we discovered two very low mass companions to 
HD\,65216, an early-M companion to HD\,177830, and we resolved the previously 
known companion to HD\,196050 into a close pair of M dwarfs. Our data
additionally confirm the bound nature of the companions to HD\,142, HD\,16141, 
and HD\,46375. The remaining 11 true companions and the two likely bound 
objects all orbit control stars. These objects are either late-K stars or 
M dwarfs, and have projected separations between 7 and 505 AU.

\subsection{The impact of stellar duplicity on the occurrence of giant planets}

A potentially sensitive point in estimating the impact of stellar duplicity on 
the occurrence of circumstellar planets is the exact definition of the control 
subsample. The
main issue is that a small amplitude radial-velocity drift can just as well
be the signature of a planet as that of a more distant stellar
companion. By being too severe on the selection of non-planet-host stars we
may thus also exclude the closest binaries from the control subsample and  
bias the analysis. To test the sensitivity of our results to the 
definition of each subsample, we performed a first analysis based on two 
different sample redefinitions: (i) a loose redefinition where both subsamples 
were little modified except for an homogeneous cut-off at close separation
($\sim$$0.7^{\prime\prime}$); (ii) a more stringent redefinition where both 
subsamples were limited in distance to 50~pc, and where control stars
showing any type of radial-velocity variation were rejected (see
\citeauthor{Eggenberger08}~\citeyear{Eggenberger08} for further details). 

According to our data, the binary fraction 
among planet-host stars is $5.5$\,$\pm$\,$2.7$\% (4/73) for the full subsample 
and $4.9$\,$\pm$\,$2.7$\% (3/62) for the redefined subsample. For control 
stars, we obtain binary fractions of $13.7$\,$\pm$\,$4.2$\% (9/66) and 
$17.4$\,$\pm$\,$5.2$\% (9/52) for the full and redefined subsamples, 
respectively. These results translate into a difference in binary fraction 
(control\,$-$\,planet-host) of $8.2\pm5.0$\% for the full subsample and of 
$12.5\pm5.9$\% for the redefined one. Although the relative errors on 
these results are quite
large due to the small number of available companions, both sample
definitions yield a positive difference with a statistical significance of
1.6-2.1$\sigma$. This positive difference means that circumstellar giant 
planets are less frequent in binaries with mean semimajor axes between 35 and 
250 AU than around single stars.

\begin{figure}
\centering
\resizebox{\textwidth}{!}{
\includegraphics{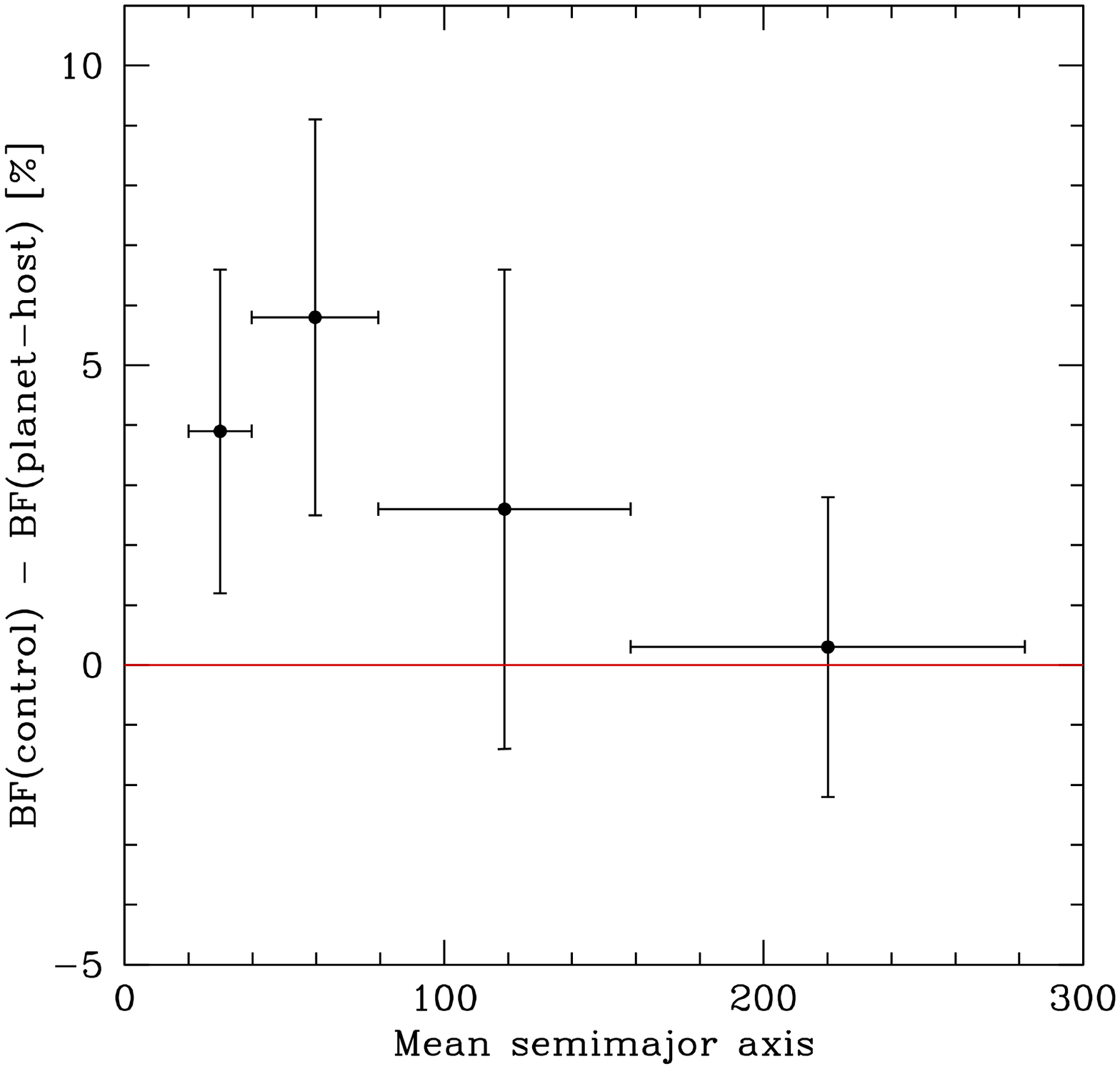}\hspace{0.1cm}
\includegraphics{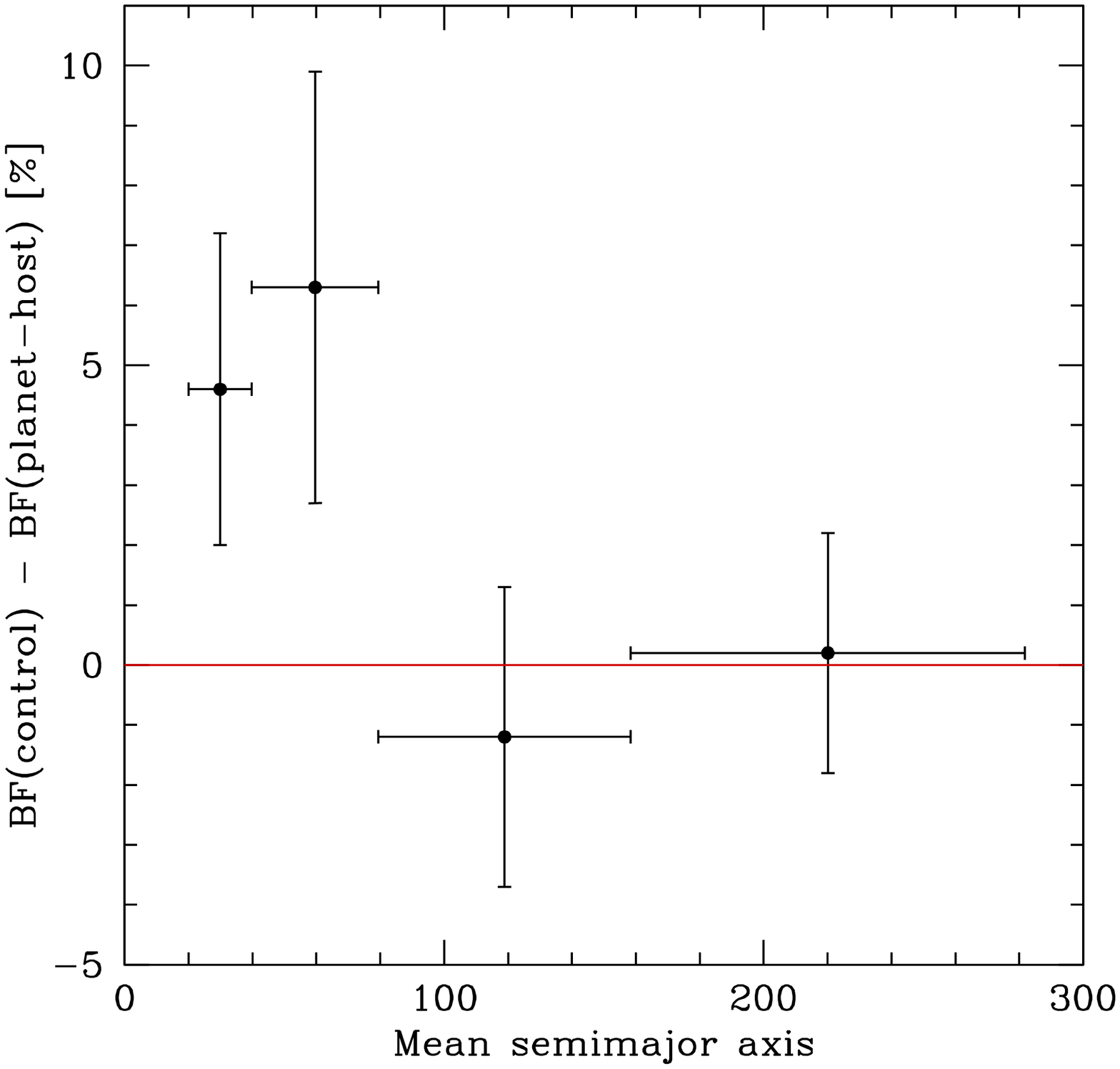}}
\caption{Difference (in percent) between the binary fraction among control stars
and the binary fraction among planet-host stars vs binary mean
semimajor axis. The left plot is based on the redefined subsamples, while the 
right plot is based on the full subsamples.}
\label{stat_res}
\end{figure}

To push the investigation a step further and to seek for a possible trend 
with mean semimajor axis, we computed the difference in binary fraction for
a few bins in separation between 20 and 280 AU. The results for both 
subsamples are displayed in Fig.~\ref{stat_res}. These two
plots show that the difference in binary fraction is not spread uniformly over
the semimajor axis range studied, but seems concentrated below $\sim$100
AU. This result is very interesting because it may corroborate the theoretical 
studies which predict a negative impact of stellar duplicity on the formation of
circumstellar giant planets in binaries $\lesssim$100 AU. At any rate, the 
statistical analysis presented here goes beyond what has been done so far, as 
the former analyses (\citeauthor{Patience02}~\citeyear{Patience02}; 
\citeauthor{Raghavan06}~\citeyear{Raghavan06}; 
\citeauthor{Bonavita07}~\citeyear{Bonavita07}; see also the contribution by 
Bonavita) could not correct their results for the selection effects of Doppler 
surveys against moderately close binaries.


\section{Results from our Doppler planet searches in spectroscopic binaries}
\label{pib}

Doppler data of binaries $\lesssim$2--6$^{\prime\prime}$ consist generally 
not of a single stellar spectrum, but of a composite spectrum made of two 
stellar spectra. Obviously, this introduces some complications into the 
extraction of the radial velocity, rendering classical cross-correlation 
techniques not well adapted to search for planets in (moderately) close systems 
(\citeauthor{EggenbergerUdry07}~\citeyear{EggenbergerUdry07}; see also the 
contribution by Marmier). A better way to extract 
precise radial velocities for the individual components of spectroscopic 
binaries is to generalize the concept of one-dimensional cross-correlation to 
that of two-dimensional correlation. This approach was followed some time ago 
by \citet{Zucker94}, who developed a two-dimensional correlation algorithm 
named TODCOR. This algorithm has recently been generalized to multiorder 
spectra \citep{Zucker03a,Zucker03b} and we are now using it to 
search for planets in spectroscopic and moderately close visual binaries. 

We present in this section some results from our ongoing searches for 
circumstellar planets 
in spectroscopic binaries. Our presentation will follow an increasing order of 
difficulty in terms of radial-velocity extraction, starting with the easiest 
systems that are single-lined spectroscopic binaries (SB1s, where only 
the spectrum of the primary star is detected) and ending with the 
more complicated double-lined spectroscopic binaries (SB2s, where the 
spectra of both components are detected).

\subsection{Planet searches in single-lined spectroscopic binaries}
\label{sb1s}

To quantify the occurrence of circumstellar giant planets in the closest 
binaries susceptible of hosting them, we initiated in 2001 a Doppler search for 
short-period circumprimary planets in SB1s 
\citep[e.g.][]{Eggenberger03,EggenbergerUdry07}. The restriction of our survey
to SB1s was mainly motivated by the higher prospect of finding circumstellar 
giant planets in these systems than in SB2s, which have similar separations but 
more massive secondaries.

Our sample of SB1s consists of 101 systems selected on the basis of former 
CORAVEL surveys carried out to study the multiplicity among nearby G and 
K dwarfs \citep{Duquennoy91,Halbwachs03}. 
All our binaries have a period of more than 1.5 year, but not
all of them have a well-characterized orbit; the systems with the longest
orbital periods (a few to several tens of years) only show long-period drifts 
in radial velocity. 
Since CORAVEL velocities have a typical precision of 300~m\,s$^{-1}$
they cannot be used to search for planets. To this purpose we took 10 to 15 
additional high-precision radial-velocity measurements of each system with the 
ELODIE spectrograph for the northern targets, and with the 
CORALIE spectrograph for the southern systems.

\subsubsection{First analysis based on one-dimensional cross-correlation}

When searching for circumstellar planets in spectroscopic binaries we are 
interested in 
short-period variations not in the radial velocities themselves but in the 
residual (radial) velocities around the binary orbits. In practice we quantify 
these variations 
by a normalized root-mean-square (rms), which is the ratio of the external 
error (i.e. the standard deviation around the orbit or around the drift) to the 
mean internal error (i.e. the mean of individual photon-noise errors). 
According to our data, most of our targets (74\%) have a normalized rms close 
to 1, which indicates that no source of radial-velocity variation other than the 
orbital motion is present. However, 12.5\% of our binaries are clearly variable 
(normalized $\mathrm{rms} > 3$), while 13.5\% of them are marginally variable (normalized 
rms between 2 and 3).

In terms of planetary prospects, the most interesting systems are the 
variable and marginally variable binaries. Yet, the presence of a circumprimary 
planet is not the only way to produce residual-velocity variations like those 
observed. Alternative possibilities include: (i) the primary star is 
intrinsically variable, (ii) the system is an unrecognized SB2 (i.e. an SB1 
when analyzed with one-dimensional cross-correlation, but an SB2 when analyzed 
with two-dimensional correlation), and (iii) the system is in fact triple and 
the secondary is itself a short-period spectroscopic binary. Assuming that 
planets are as common in moderately close binaries as around single stars, we 
expect to find $\sim$2 planets more massive than 0.5~M$_{\rm Jup}$ and 
with a period $\lesssim$40~days in our sample. This estimation 
shows that most of the observed residual-velocity variations likely stem from 
the binary or multiple nature of our targets. To disentangle the few potential
planet-host stars from the unrecognized SB2s and triple systems, we are 
analyzing all the variable and marginally variable systems with the TODCOR 
algorithm. 

\begin{figure}
\centering
\resizebox{\textwidth}{!}{
\includegraphics{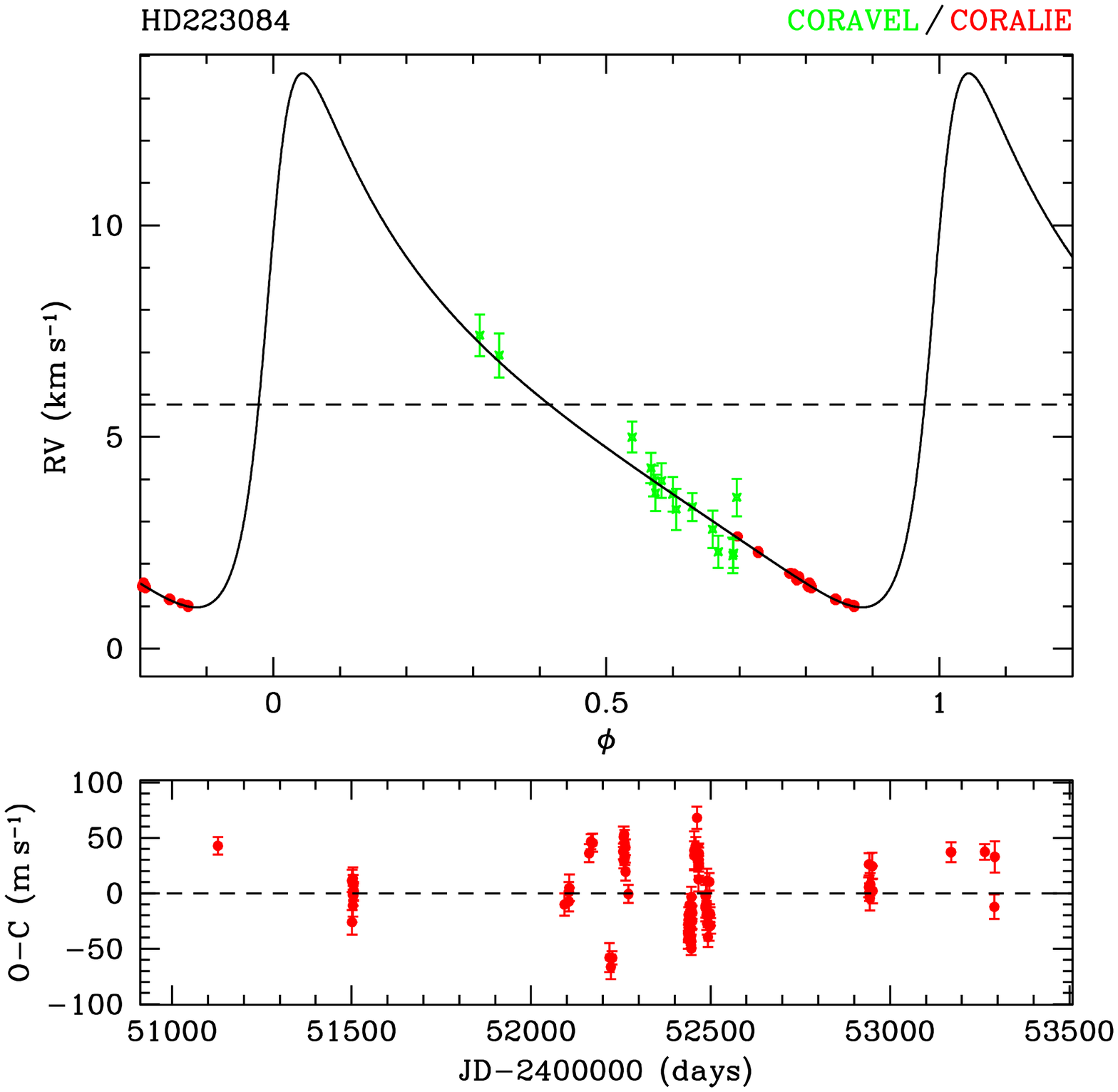}
\includegraphics{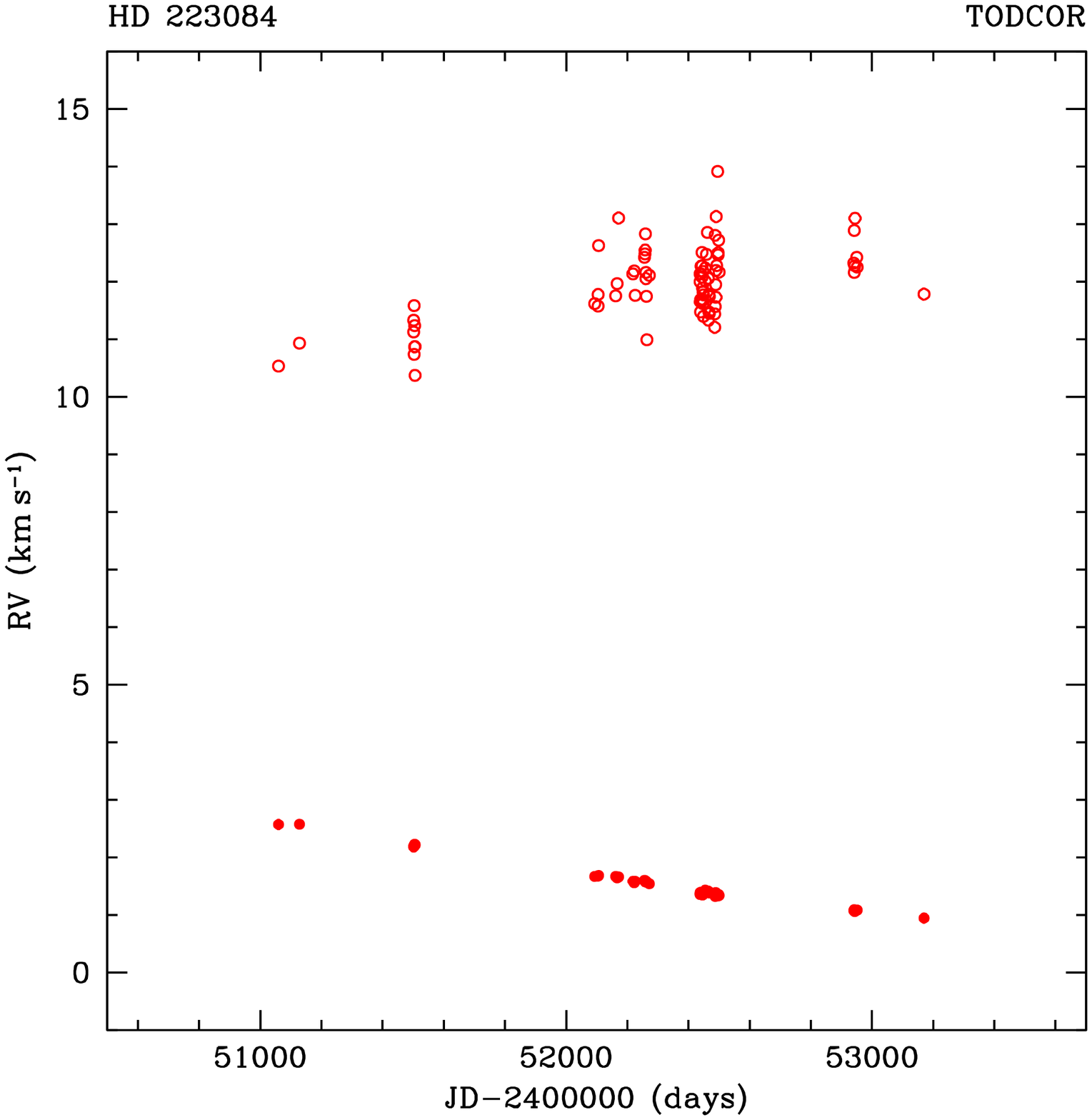}}
\resizebox{0.5\textwidth}{!}{
\includegraphics{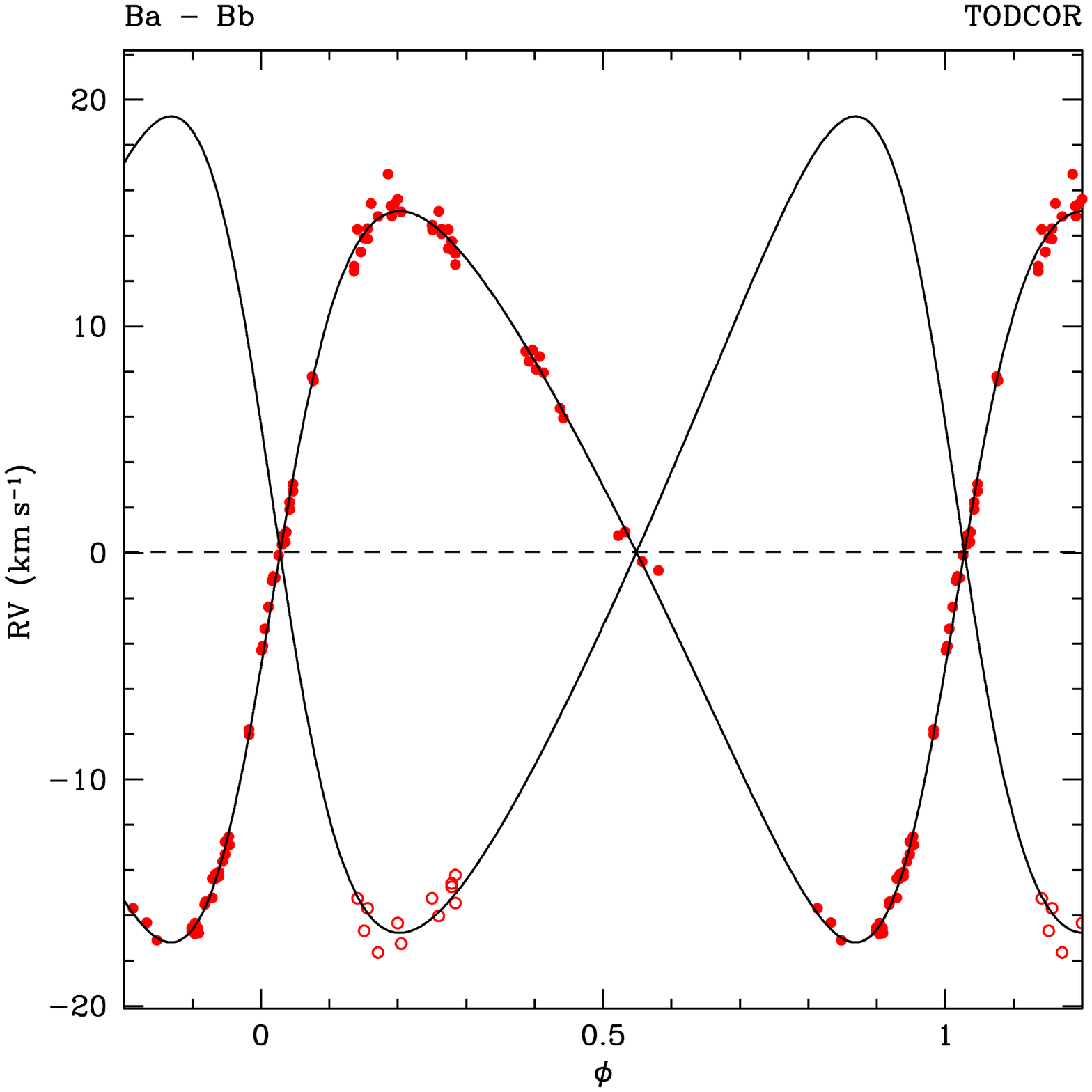}}
\caption{An example of triple system: HD\,223084. 
{\bf Top left:} CORAVEL (crosses, large error bars) and CORALIE 
(dots) velocities for HD\,223084. The binary orbit is tentative and is used 
only as a proxy to compute residual velocities. The bottom panel shows the
residual velocities (CORALIE data only). 
{\bf Top right:} TODCOR velocities for HD\,223084\,A (dots) and  
HD\,223084\,Ba (open circles) after having removed the 202-day modulation of
the Ba--Bb inner pair. 
{\bf Bottom:} SB2 orbit for  HD\,223084\,Ba (dots) and HD\,223084\,Bb
(open circles). This orbit is characterized by a period of 202~days.}
\label{hd223084}
\end{figure}

\subsubsection{Complementary analysis based on two-dimensional correlation}

Among the four variable binaries studied in detail so far, two turned out to be 
triple systems (see Fig.~\ref{hd223084} for an example) and the two others 
turned out to be unrecognized SB2s. None of these systems shows hints of 
the presence of a circumprimary planet.

\subsubsection{Preliminary statistical results}

In 74\% of our SB1s the secondary component is so faint (magnitude 
difference $\Delta V$\,$\gtrsim$\,$6$) that it does not contribute 
significantly to the recorded flux. The precision achieved on the measurement 
of the radial velocity of the primary star is then as good as for single stars. 
In 26\% of our binaries, the secondary component contributes to some extent 
to the recorded flux ($\Delta V$\,$\in$\,$[\sim 3,\sim 6]$), rendering the use 
of two-dimensional correlation mandatory to search for circumprimary planets. 
For these systems, we estimate that typical precisions on the radial velocity 
of the primary star range from 10 to 20~m\,s$^{-1}$. Although these precisions 
are not as good as for single stars, they remain good enough to search for 
giant planets. 

So far our survey has unveiled no promising planetary candidate, but the data 
of 22 variable and marginally variable systems remain to be analyzed in detail 
with two-dimensional correlation. Since contamination effects stemming 
from the stellar companions are likely to prevail over potential planetary 
signals, two-dimensional analyses must be completed before concluding on the 
existence -- or absence -- of planets in our sample. All we can say at
present is that less than 22\% of the SB1s from our sample host a short-period 
($P$\,$\lesssim$\,40~days) giant (M$_2\,\sin{i} \gtrsim 0.5$~M$_{\rm Jup}$)
circumprimary planet. The final analysis will provide a much stronger 
constraint on the frequency of short-period giant planets in SB1s.

\subsection{Planet searches in double-lined spectroscopic binaries}
\label{sb2s}

SB2s have not been systematically included in any of our programs yet, but we 
are conducting a series of observational tests to characterize the feasibility 
of Doppler planet searches in and around these systems. 
As an illustration of this work, we present here the results we have obtained 
for our best-studied case, the triple system HD\,188753 \citep{Eggenberger07a}.

\subsubsection{The example of HD\,188753}

HD\,188753 has attracted much attention since \citet{Konacki05a} 
reported the discovery of a Jovian planet on a 3.35-day orbit around the 
primary component of this triple system. Aside from the planet, HD\,188753 
consists of a visual pair (HD\,188753\,A and HD\,188753\,B) with 
a semimajor axis of 12.3~AU ($0.27^{\prime\prime}$\ separation) and 
an eccentricity of 0.5 \citep{Soederhjelm99}. In addition, component B is itself 
a spectroscopic binary (i.e. HD\,188753\,B is composed of HD\,188753\,Ba and 
HD\,188753\,Bb) with a period of 155~days \citep{Griffin77,Konacki05a}. 
What renders this planet discovery particularly noteworthy is that according to 
the canonical models of planet formation the periastron distance of the AB pair 
may be small enough to preclude giant planet formation around HD\,188753\,A 
\citep{Nelson00,Mayer05,Boss06,Jang-Condell07}. The discovery of a close-in 
giant planet around this star has thus been perceived as a serious challenge to
planet-formation theories, though the alternative possibility that 
HD\,188753\,A may have acquired its planet through dynamical interactions was
also pointed out \citep{Pfahl05,PortegiesZwart05}.

\begin{figure}
\centering
\resizebox{\textwidth}{!}{
\includegraphics{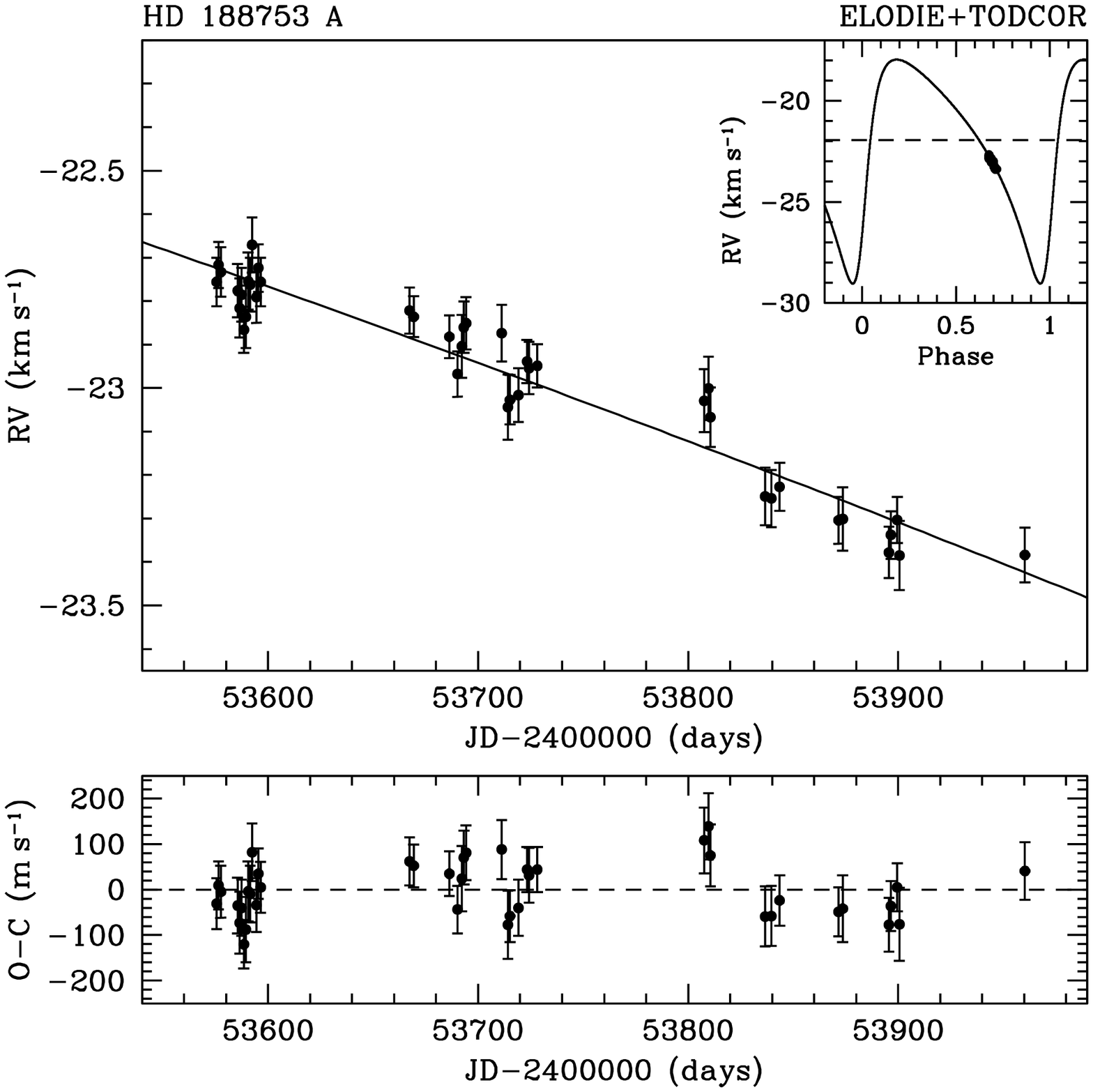}
\includegraphics{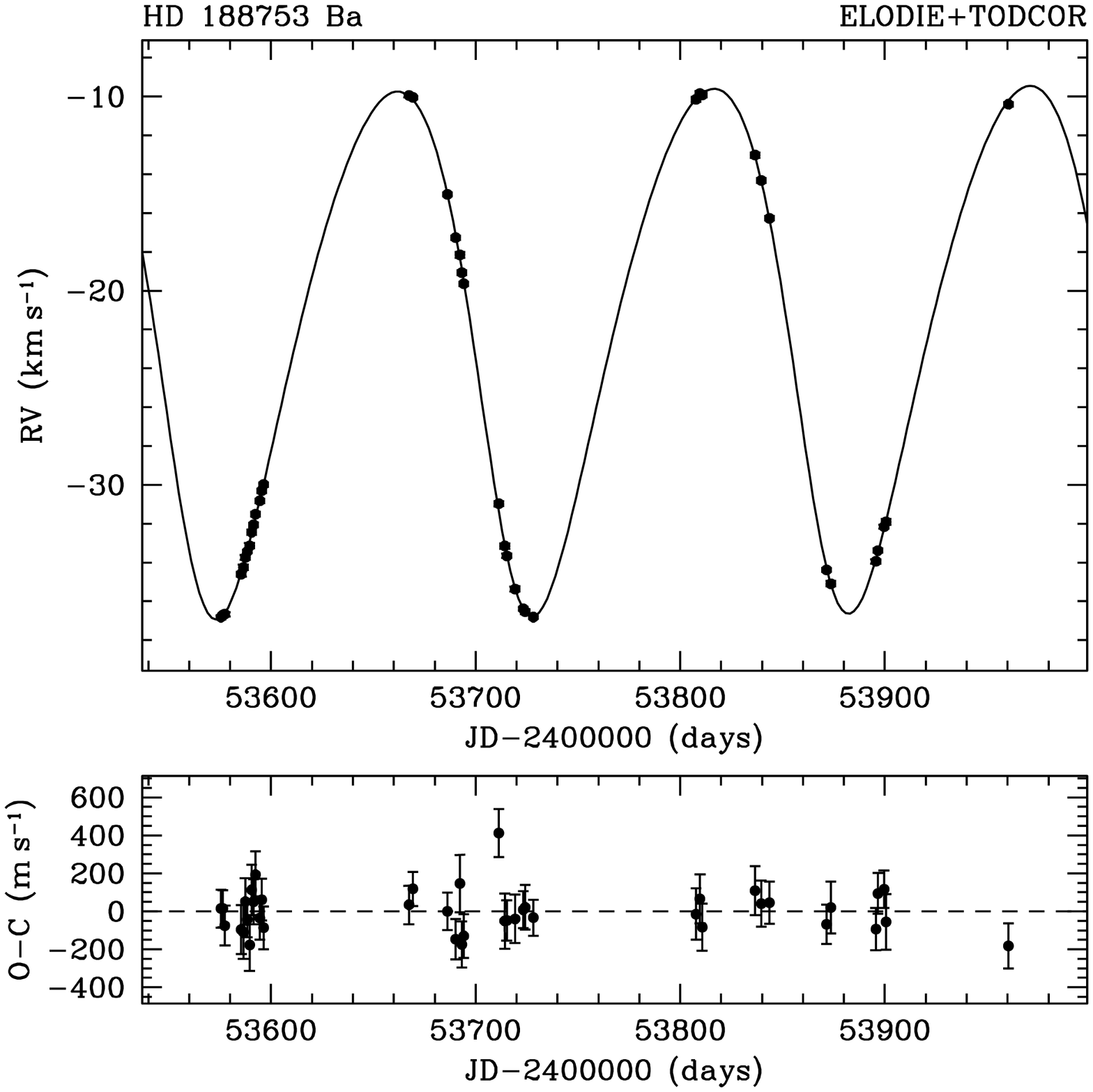}}
\caption{Radial velocities and orbital solutions for 
HD\,188753\,A (left) and HD\,188753\,Ba (right). For component A, the solid 
line represents the 25.7-year orbital motion of the visual pair shown in
full in the inset. For component Ba, the orbital solution corresponds to the
155-day modulation and it includes a linear drift to take the 25.7-year orbital 
motion into account.}
\label{hd188753}
\end{figure}

Following the discovery announcement, we monitored 
HD\,188753 with the ELODIE spectrograph during one year. Our TODCOR velocities 
for the two brightest components -- HD\,188753\,A and HD\,188753\,Ba -- are 
displayed in Fig.~\ref{hd188753}. Using two-dimensional correlation, the 
spectrum of the faintest component is undetectable in most 
of our observations. Our measurements confirm that HD\,188753\,Ba is a 
spectroscopic binary with a period of 155~days. However, our radial velocities 
for HD\,188753\,A show no sign of the 3.35-day planetary signal reported by 
\citet{Konacki05a}. Instead, the residuals around the velocity drift due to the 
orbital motion of the AB pair are basically noise, and the rms of 
60~m\,s$^{-1}$ can be interpreted as the precision we achieve on the 
measurement of the radial velocity of this star. 
Monte Carlo simulations showed that we had both the precision and the 
temporal sampling required to detect a planetary signal like the one described 
by \citet{Konacki05a}. On that basis, we conclude that our ELODIE data show no 
evidence of a 1.14-M$_{\rm Jup}$ planet on a 3.35-day orbit around 
HD\,188753\,A.

Besides the question of whether there is a hot Jupiter around HD\,188753\,A, 
our analysis of HD\,188753 raises several issues. 
In particular, the precision of 60~m\,s$^{-1}$ obtained on the radial 
velocity of HD\,188753\,A looks abnormally poor compared to the results 
presented in Sect.~\ref{sb1s}. The most probable explanation is that the search 
for circumprimary planets in SB2s requires higher quality data (mainly a better 
spectral resolution) than the search for circumprimary planets in SB1s. The 
new data set of HD\,188753 we have just acquired with the SOPHIE spectrograph
(spectral resolution of 75,000 against 40,000 for ELODIE) will help 
clarify this point. But these new data will not likely change our 
conclusion that there is no supporting evidence for the claimed hot Jupiter 
around HD\,188753\,A.

\subsubsection{Outlook}

Including SB2s in Doppler planet searches is desirable for two reasons. 
Firstly, the frequency of circumstellar giant planets residing in these systems 
would provide important constraints for planet formation theories. 
Secondly, some SB2s are the potential targets for circumbinary 
planet searches, which constitute a still largely unexplored research field 
worth of interest. As illustrated above, deriving radial velocities for the
individual components of SB2s to the precision needed to search for planets is 
challenging. A few different methods are currently being tested to determine 
the most efficient way to overcome this challenge and the main limiting factors 
associated with each method. 
Current results indicate that the prospects of using Doppler spectroscopy to 
search for giant planets in SB2s are promising, provided one has good enough 
data (see e.g. the contribution by Konacki).


\section{Conclusion and perspectives}
\label{conclusion}

Over the past seven years, binaries have become increasingly interesting targets
for planet searches. From the observational perspective, Doppler surveys have 
shown that at least 17\% of the known planetary systems are associated 
with one or more stellar companions \citep[e.g.][]{EggenbergerUdry07} 
and that circumstellar giant planets exist in binaries as close as 20 AU 
\citep{Queloz00,Hatzes03,Zucker04,Correia08}. From the theoretical 
perspective, simulations indicate that the presence of a stellar companion 
within $\sim$100~AU likely affects the formation and evolution of circumstellar 
giant planets \citep{Kley00,Nelson00,Mayer05,Boss06,Thebault06}, 
leaving potential imprints in the occurrence and properties of 
these objects. Circumstellar planets residing in binaries $\lesssim$100~AU 
thus provide unique testing grounds for the models of planet formation and 
evolution. 

Imaging and literature surveys searching for stellar companions to the known 
planet-bearing stars have been very successful, revealing many new binary and 
multiple planet-host systems 
\citep[e.g.][]{Patience02,Raghavan06,Chauvin06,Desidera07,Eggenberger07b,Mugrauer07}. 
Thanks to its well-defined control sample, our NACO survey provides the first 
piece of evidence that circumstellar giant planets are intrinsically less 
frequent in binaries $\lesssim$100~AU than around single stars 
\citep{Eggenberger08}. Future analyses based on larger samples will constrain
the dependence of this frequency on binary separation over the range
$\sim$35-250 AU.

The discoveries from classical Doppler planet searches and from imaging surveys 
indicate that the Kozai mechanism plays a role in shaping the high end of the 
eccentricity distribution of extrasolar planets 
\citep{Wu03,Takeda05,Tamuz08,Moutou09}. Additionally, the distinctive 
properties of the short-period planets residing in binary or hierarchical 
systems \citep{Zucker02,Eggenberger04,Desidera07,Mugrauer07} suggest that some 
of these planets owe their current orbital configuration to Kozai migration 
\citep{Takeda06,Fabrycky07,Wu07}. The data set on planets in binaries 
thus provides growing evidence that distant stellar companions commonly affect 
the orbital evolution of planetary systems on secular timescales.

Since 2002 significant efforts have been put into extending planet searches to 
(moderately) close binaries using different techniques, including Doppler 
spectroscopy, phase-referenced interferometry, eclipse or pulse timing, transit 
photometry, gravitational microlensing, and adaptive optics imaging. Although 
these efforts have not produced many planet discoveries yet, they will 
yield very important results in the coming years. Upon completion, the ongoing
Doppler and interferometric surveys will provide the first measures of the
true frequency of circumstellar planets in binaries $\lesssim$50 AU. When
combined with the results from imaging programs, these new data will give 
some information as to whether circumstellar giant planets commonly form in 
binaries $\lesssim$100~AU. At the same time, the present and future Doppler 
surveys targeting SB2s will quantify the frequency of massive circumbinary 
planets.

In the next few years, several additional observing facilities 
will open up new possibilities for planet searches in/around (moderately) close 
binaries. Thanks to their high photometric precision and continuous monitoring 
of rich stellar fields, the recently launched 
CoRoT\footnote{http://smsc.cnes.fr/COROT/} 
and Kepler\footnote{http://kepler.nasa.gov} 
missions will initiate large-scale planet searches around eclipsing binaries  
(\citeauthor{Doyle04}~\citeyear{Doyle04}; 
\citeauthor{Ofir09}~\citeyear{Ofir09}; see also the contribution by 
Sybilski). Alternatively, new interferometric facilities like PRIMA 
at the Very Large Telescope Interferometer \citep{Delplancke08} or the ASTRA 
upgrade of the Keck Interferometer \citep{Pott08} will allow to extend 
astrometric planet searches to significantly wider and fainter binaries. 
The upcoming generation of high-contrast adaptive optics instruments such as 
HiCIAO at Subaru \citep{Tamura06}, GPI at Gemini South \citep{Macintosh08},  
SPHERE at the Very Large Telescope \citep{Beuzit08} and PALM-3000/Project-1640 
at Palomar \citep{Bouchez08} will also likely be used to 
carry out systematic searches for giant planets in some types of moderately 
close binaries. Gathering together the results from all these different surveys 
we will then have pretty good answers to the two questions mentioned at the 
beginning of this chapter: What types of binaries do host planets 
in S/P/L-type orbits?, and Are such planets common or rare?

As surveys progress and diversify, the conviction that planets are common 
objects in the universe continually strengthen. In addition to the encouraging 
observational results obtained so far on circumstellar giant planets, 
theoretical studies support the existence of low-mass planets in many 
binary systems (e.g. \citeauthor{Thebault06}~\citeyear{Thebault06};
\citeauthor{Haghighipour07}~\citeyear{Haghighipour07};
\citeauthor{Quintana07}~\citeyear{Quintana07}; see also the contribution by 
Marzari). Similarly, circumbinary planets are expected to be
quite common \citep[e.g.][]{Quintana06,Pierens07,Scholl07,Pierens08a}. 
Observational programs targeting (moderately) close binaries thus promise
additional exciting results to come. Since a full understanding of planet 
formation must address the issue of circumstellar and circumbinary planets, 
these programs are an essential part of today's research on extrasolar planets.



\end{document}